\documentclass[showpacs,preprintnumbers,amsmath,amssymb,prb,twocolumn,superscriptaddress,floatfix]{revtex4}
\usepackage{graphicx}
\usepackage{bm}

\newcommand{\f}[1]{^{(#1)}} 
\newcommand{\ve}[1]{\mathbf{#1}}

\newcommand{\im}{{\rm i}}
\newcommand{\eq}[1]{Eq. (\ref{#1})}
\newcommand{\xy}{3D$xy$\ }

\newcommand{\fig}[1]{Fig. \ref{#1}}
\newcommand{\beg}{\begin{eqnarray}}
\newcommand{\eee}{\end{eqnarray}}
\def\comment#1{}

\begin{document}

\title{Observation of a metallic superfluid in a numerical experiment}

\author{E. Sm{\o}rgrav}
\affiliation{Department of Physics, Norwegian University of Science and Technology, N-7491 Trondheim, Norway}

\author{E. Babaev}
\affiliation{Laboratory of Atomic and Solid State Physics, Cornell University, Ithaca, NY 14853-2501, USA}
\affiliation{Department of Physics, Norwegian University of Science and Technology, N-7491 Trondheim, Norway}

\author{J. Smiseth}
\affiliation{Department of Physics, Norwegian University of Science and Technology, N-7491 Trondheim, Norway}

\author{A. Sudb{\o}}
\affiliation{Department of Physics, Norwegian University of Science and Technology, N-7491 Trondheim, Norway}


\begin{abstract}
We report the observation, in Monte Carlo
simulations, of a novel type of quantum ordered state: {\it the metallic
superfluid}. The metallic superfluid features ohmic resistance 
to counter-flows of protons and electrons, while featuring dissipationless 
co-flows of electrons and protons. One of the candidates for a physical realization 
of this remarkable state of matter is hydrogen or its isotopes under high compression. 
This adds another potential candidate to the presently known quantum dissipationless
states, namely superconductors, superfluid liquids and vapours, and
supersolids.  
\end{abstract}

\pacs{71.10.Hf, 74.10.+v, 74.90.+n,11.15.Ha} 

\maketitle

  At low temperatures, fluids become dominated by the wavelike nature of their constituent 
  particles when the thermal de Broglie wavelength exceeds the interparticle separation. Such 
  quantum fluids usually feature superconductivity or superfluidity, which however may be 
  destroyed by topological line-defects (vortices) threading the entire system.  Vortices may 
  be induced by a magnetic field or rotation \cite{Abrikosov}, or by thermally excited
  {\it transverse} phase fluctuations of the macroscopic wavefunction of superconductors and 
  superfluids \cite{ZBT,AKAS,Fossheim_Sudbo_book}. In a system of rapidly growing interest, a 
  two-component superconductor \cite{BSA}, vortices yield dramatic physical consequences. Here 
  we report the first observations, in a numerical experiment, of a novel type of quantum fluid 
  originating in aggregate states of vortex matter. Increasing temperature, a composite vortex 
  lattice melts into a composite vortex liquid, whence superconductivity is lost while superfluidity 
  is retained, {\it yielding a metallic (ohmic) superfluid}. At higher temperature, another unusual transition 
  occurs where the composite vortex liquid  ``ionizes" into a  ``plasma" of constituent vortices, 
  destroying superfluidity.
  
  Recently, there has been considerable interest in theories of superconductors with several 
  superconducting components coupled by a magnetic field, but with no possibility of Josephson 
  tunneling of one component into another. This is predicted to occur in a wide variety of 
  physical systems, most notably in condensed matter such as hydrogenic atoms subjected to extreme 
  pressure \cite{ja,egor2002,frac,BSA,smiseth2004,smorgrav2005}
  or effective theories of easy-plane quantum antiferromagnets \cite{senthil2003}. Renewed interest 
  in the long sought liquid metallic hydrogen (LMH) is due to recent ab initio calculations
  \cite{Bonev} along with a breakthrough in synthesis of ultrahard artificial diamonds, essential 
  for obtaining the required extreme pressures in anvil cells \cite{newlink}. These facts, along 
  with a recent measurement of an unusually low melting temperature of dense Na \cite{Hemley2005}, 
  strongly hint at a realization of LMH in the not too distant future. Thus, understanding the 
  superconducting properties of this system is important, since magnetic field 
  experiments can be conducted in high pressure anvil cells, possibly confirming the realization 
  of this novel state of matter in a terrestrial laboratory. It is commonly 
  accepted that LMH is abundant in the interior of Jupiter and Saturn and quite possibly  
  also present in some of the known $200$ extrasolar giant planets \cite{Guillot2004}. In these
  cases, however, LMH is conjectured to exist in the classical metallic regime at several
  thousand degrees Kelvin. In contrast, the state projected to exist in 
  Refs. \onlinecite{ja} is a ground state quantum fluid, and is 
  the one we focus on in this Letter.

  Realization of LMH could well constitute the next milestone in quantum 
  fluids. It is projected to feature Cooper-pairs of both electrons and protons at low temperatures    
  \cite{ja}. The resulting quantum fluid differs radically from previously 
  known quantum fluids, in that its physical properties cannot be classified exclusively as a 
  superconductor or a superfluid \cite{BSA}.  Remarkably, such a system features both 
  superconductivity and superfluidity which appear as collective phenomena corresponding 
  to co- and counter-flows of two species of Cooper-pairs, with a complicated interplay 
  between them. Thus, LMH has been  conjectured to sustain phase transitions connecting a 
  superconducting {\it and} superfluid state to a metallic state featuring a superfluid 
  mode \cite{BSA}, or to a superconducting state with no superfluid mode.  
  
  The transition from a state featuring superconductivity {\it and} superfluidity to a state where 
  superfluidity is lost and superconductivity is retained, has recently been observed in a large 
  scale Monte Carlo (MC) simulation \cite{smorgrav2005}. However, the remarkable possibility of 
  a transition from a ``composite"  vortex lattice (superconducting state where vortex matter 
  forms a solid) to a ``composite vortex liquid" (metallic superfluid state of the system) along 
  with a subsequent transition from a composite vortex liquid to ``vortex line plasma", {\it has 
  thus far not been confirmed}. {\it It is the purpose of this Letter to report an observation of 
  the metallic superfluid, not dealt with in simulations previously}, in a numerical experiment.

  The superconducting phase of LMH is given by the Ginzburg-Landau model with two scalar 
  fields  $\Psi_0^{(1)}(\ve r)$ and $\Psi_0^{(2)}(\ve r)$ describing superconducting
  condensates of protons and electrons, respectively. It is
  defined via the energy density
  \begin{eqnarray}
    \label{gl_action}
    {\cal{H}} = \sum_{\alpha=1}^2 \frac{|\ve D \Psi_0^{(\alpha)}(\ve r)|^2}{2 M^{(\alpha)}}
    +  V(\{|\Psi_0^{(\alpha)}(\ve r)|\}) + \frac{1}{2}(\nabla\times\ve{A}(\ve r))^2.
  \end{eqnarray}
  Here, $M^{(1)}$ and $M^{(2)}$ are the masses of the condensates, the covariant derivative is given
  by $\ve D \equiv \nabla -\im e\ve{A}(\ve r)$, and $V(\{|\Psi_0^{(\alpha)}(\ve r)|\})$ is the potential 
  term.  The particular form of  $V(\{|\Psi_0^{(\alpha)}(\ve r)|\})$ is not essential for the 
  large-scale physics we address, but its dependence only on $|\Psi^{(\alpha)}_0|$  reflects the fact 
  that Cooper-pairs of electrons cannot be converted into Cooper-pairs of protons, and vice versa. For 
  the issues discussed in this paper, it suffices to work in the phase only approximation   
  $\Psi_0^{(\alpha)}(\ve r) =|\Psi_0^{(\alpha)}|\exp [ \im\theta^{(\alpha)}(\ve r)]$ where
  $|\Psi^{(\alpha)}_0|$ is treated as a constant \cite{smorgrav2005}. In LMH, the two order 
  parameters correspond to electronic and protonic Cooper-pairs. Moreover, \eq{gl_action} may 
  be rewritten as follows  \cite{egor2002,frac,BSA,smorgrav2005}
  \begin{eqnarray} 
    \label{charge_neutral}
    {\cal H} & = & \frac{1}{2 \Psi^2} ~ \left( |\psi^{(1)}|^2 \nabla
      \theta^{(1)}
      +|\psi^{(2)}|^2 \nabla \theta^{(2)}  - e \Psi^2 \ve A \right)^2\\
    & + & \frac{1}{4 \Psi^2} |\psi^{(1)}|^2 ~ |\psi^{(2)}|^2 ~
    \left( \nabla (\theta^{(1)}- \theta^{(2)}) \right)^2 + \frac{1}{2}(\nabla\times\ve{A})^2,\nonumber
  \end{eqnarray}
  where $|\psi^{(\alpha)}|^2 = |\Psi^{(\alpha)}_0|^2/M^{(\alpha)}$, and
  $\Psi^2 \equiv |\psi^{(1)}|^2+|\psi^{(2)}|^2$. The neutral and charged modes, 
  described by the second and first term in Eq. (\ref{charge_neutral}), are 
  explicitly identified.   
  
  The topological objects of \eq{gl_action} are vortices of type-$1$ and 
  type-$2$ defined by a $2\pi$ winding in $\theta^{(1)}$ and $\theta^{(2)}$, 
  respectively. The interaction potential between these vortices is a 
  superposition of a Coulomb potential and a Yukawa potential, arising 
  out of the neutral and the charged mode, respectively. (For a  
  derivation of the formulae for the vortex interaction, see Ref.
  \onlinecite{smorgrav2005}). The energy of a vortex associated with $\pm 2 \pi$ 
  winding in $\theta^{(1)} - \theta^{(2)}$ is logarithmically divergent. As the 
  neutral mode tends to lock 
  $\nabla\theta^{(1)}$ to $\nabla\theta^{(2)}$, in order to minimize the 
  second term in Eq. (\ref{charge_neutral}), the vortices of type-$1$ and 
  type-$2$ pair up into a composite vortex for which 
  $\nabla(\theta^{(1)}-\theta^{(2)})=0$. Therefore, a composite vortex
  is an object where a type-$1$ and a type-$2$ vortex are co-centered
  and co-directed in space, and the Coulomb part of
  the pair-potential exactly cancels. The screened potentials add, but 
  the associated overall energy is finite
  \cite{frac,BSA,smiseth2004,smorgrav2005}. 
  In the presence of an externally applied magnetic field, the ground state
  of the system is a lattice of co-centered vortices of type-$1$ and type-$2$,
  {\it a composite vortex lattice}.

We have performed MC simulations on \eq{gl_action} at finite temperature using local Metropolis 
updating on the fields $\theta^{(1)}(\ve r)$, $\theta^{(2)}(\ve r)$, and $A(\ve r)$. The 
system size we have used is $L \times L \times L$, with $L = 120$. The coupling 
constants investigated are $|\psi^{(1)}|^2=0.5$, $|\psi^{(2)}|^2=1.0$ and $e=1.0$
and the external magnetic field is $\ve B=\nabla\times \ve A(\ve
r)=(0,0,2\pi f)$ with $f=1/20$. Thus, there are $20$ plaquettes in the $xy$-plane 
for each field-induced vortex. The external magnetic field is imposed 
by splitting $\ve A(\ve r)=\ve A_{F}(\ve r)+\ve A_0(\ve r)$, where
$\ve A_{F}$ is free to fluctuate subject to periodic boundary 
conditions, and $\ve A_0=(0,2\pi x f,0)$ is kept fixed. We have chosen 
the amplitude ratios $|\psi^{(2)}|^2/|\psi^{(1)}|^2=2.0$ for numerical 
convenience. We emphasize that the results in 
this paper are dictated by symmetry, and the physical picture we present 
will thus be representative also for LMH, where $|\psi^{(2)}|^2/|\psi^{(1)}|^2 \sim  10^3$. 

To find the lattice ordering of vortices we compute the planar structure function 
$S^{(\alpha)}(\ve k_\perp)$ of the local vorticity $\ve n^{(\alpha)}(\ve r)$, defined 
by $\ve{\Delta} \times [\ve{\Delta} \theta^{(\alpha)} - e \ve A] = 2 \pi \ve n^{(\alpha)}(\ve r)$, 
given by
$S^{(\alpha)}(\ve k_\perp) = 
\langle | \sum_{\ve r} ~ n_z \f{\alpha}(\ve r) ~ e^{i \ve k_\perp \cdot \ve r_\perp} |^2 \rangle/(f L^3)^2.$
Here, $\Delta_\mu$ is the lattice difference operator, 
${\Delta}_\mu \theta^{(\alpha)} - e A_\mu \in [0, 2 \pi \rangle$,
$\ve r$ runs over the possible positions of the vortices, and $\ve k_\perp$ and $\ve r_\perp$ 
are perpendicular to $\ve B$. If vortices form a lattice, $S^{(\alpha)}(\ve k_\perp)$ will exhibit 
a six-fold symmetric Bragg structure, and feature a  ring-structure in 
the vortex liquid phase.

\begin{figure}[htb]
\centerline{\scalebox{0.55}{\rotatebox{-90.0}{\includegraphics{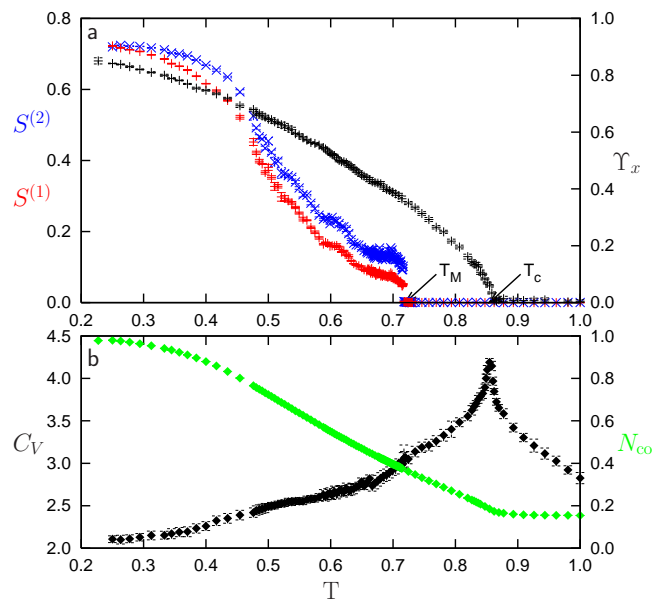}}}}
\caption{ \label{Monte_Carlo_res} (Color online)
Results of numerical experiments on a two-component vortex system, for 
$|\psi^{(1)}|^2=0.5$, $|\psi^{(2)}|^2=1.0$, and $e=1.0$. Panel {\bf a:} Structure functions 
for protonic $S^{(1)}(\ve K)$ (red) and electronic $S^{(2)}(\ve K)$(blue)  vortices, for 
$\ve K = (\pi/4, 2 \pi/5)$. They drop to zero discontinuously at $T_{M}$ where the system 
looses its superconducting properties, but retains superfluidity, as evidenced by a finite 
helicity modulus $\Upsilon_x$ (black line). $\Upsilon_x$ 
serves as an order parameter in the vortex liquid phase, dropping to zero at $T_C$. Panel {\bf b}: 
$N_{\rm co}$(green) is finite across the melting transition, but drops continuously to a 
small value at $T_C$, where it has a kink. At $T_C$, the specific heat $C_V$ has an anomaly.}
\end{figure}

The MC results are given in \fig{Monte_Carlo_res}, showing the structure 
function $S^{(1)}(\ve K)$ (red) for protonic vortices, and  
$S^{(2)}(\ve K)$ (blue) for electronic vortices, where ${\ve K}=(\pi/4,2 \pi/5)$ is  
a Bragg vector. In the low temperature regime both functions are finite, but decrease 
gradually as the temperature is increased. At $T_M$, $S^{(1)}(\ve K)$ and $S^{(2)}(\ve K)$ 
{\it vanish at the same temperature even though the ratio of the bare stiffnesses of the 
condensates is $2.0$}. Moreover, both structure functions vanish 
discontinuously, the hallmark of a first order melting transition of the 
{\it composite} vortex lattice. 

To probe the splitting of the co-centered vortices into constituent vortices, we compute 
the vortex co-centricity, defined as   
$N_{\rm{co}} \equiv N_{\rm{co}}^{(+)} - N_{\rm{co}}^{(-)}$
where 
$N_{\rm{co}}^{(\pm)} = 
\sum_{\ve r} |n_z \f 2(\ve r)| \delta_{n_z \f 1(\ve r),\pm n_z \f 2(\ve r)}/\sum_{\ve r}|n_z \f 2(\ve r)|$,
and $\delta_{i,j}$ is unity if $i=j$ and zero otherwise. Therefore, $N_{\rm{co}}$ is the 
fraction of type-$2$ vortex segments that are co-centered and co-linear with type-$1$ segments. 
We find that passing through the first-order melting transition renders  $N_{\rm{co}}$ unaffected, 
whence we conclude that the observed transition is a melting of a {\it composite} vortex lattice. 

In one-component type-II superconductors, melting of the vortex lattice amounts to a complete 
destruction of dissipationless currents \cite{Fossheim_Sudbo_book}. To follow the 
fate of the superfluid mode of the two-component system in the vortex liquid state, we measure 
the ordering in $\gamma(\ve r) \equiv \theta^{(1)}(\ve r)-\theta^{(2)}(\ve r)$. 
To probe the global phase coherence in this variable we consider the helicity modulus $\Upsilon_\mu$, 
equivalently the superfluid density, 
given by 
$\Upsilon_\mu = \left[ \left\langle c \right\rangle - \eta ~ \left\langle s^2 \right\rangle \right]/L^3$,
where 
$c=\sum_{\ve r}\cos{(\Delta_\mu \gamma(\ve r))}$, 
$s=\sum_{\ve r} \sin{(\Delta_\mu \gamma(\ve r))}$, 
and $\eta=\beta|\psi^{(1)}|^2|\psi^{(2)}|^2/2 \Psi^2$.
When the composite vortex lattice melts, destroying superconductivity \cite{Fossheim_Sudbo_book}, 
{\it the superfluid density remains unaffected}, cf. Fig. \ref{Monte_Carlo_res}. {\it Hence, 
this transition separates a superconducting superfluid from a metallic superfluid 
state}. 
 
Thus, we have identified a transition from a composite vortex lattice into a composite 
vortex liquid.  Increasing the temperature further,  we find a phase transition between 
the {\it composite vortex liquid} and the {\it ``ionized vortices" plasma}, as evidenced 
by the vanishing of the superfluid density $\Upsilon_x$, cf. Fig. \ref{Monte_Carlo_res}. 
{\it We observe that this transition is accompanied by a pronounced anomaly in the 
specific heat $C_V$, indicating that this is a critical phenomenon}. This is 
corroborated by the following physical argument. The liquid state of vortices in 
an ordinary superconductor is a state where translational symmetry is restored and 
superconductivity is lost \cite{Fossheim_Sudbo_book}. In the {\it composite vortex 
liquid} state, every electronic vortex is accompanied by a protonic vortex performing 
only finite excursions away from the electronic vortex line. Therefore, for every plane 
slicing an electronic vortex in a direction perpendicular to the magnetic field, it is 
possible to identify a finite length closed contour which also encompasses an accompanying 
co-directed protonic vortex. Along such a contour, there is no nontrivial winding in 
the phase-difference $\gamma$. Therefore, melting of the composite vortex lattice into 
a composite vortex liquid does not restore the broken global $U(1)$-symmetry associated 
with $\gamma$.  On the other hand, in the vortex plasma state one cannot find a protonic 
vortex accompanying every electronic vortex, implying a disordering of $\gamma$. Hence, the 
associated global $U(1)$-symmetry  is restored during the ``vortex ionization" transition 
taking place within the vortex liquid, and it is therefore in the \xy universality class.
   
To gain further insight, we have also extracted 
vortices and visualized snapshots of configurations of vortex matter 
in  small segments of the system. The results are shown in Fig. \ref{charge1}.
\begin{widetext}
\begin{center}
\begin{figure}[htb]
{\scalebox{0.65}{\rotatebox{0.0}{\includegraphics{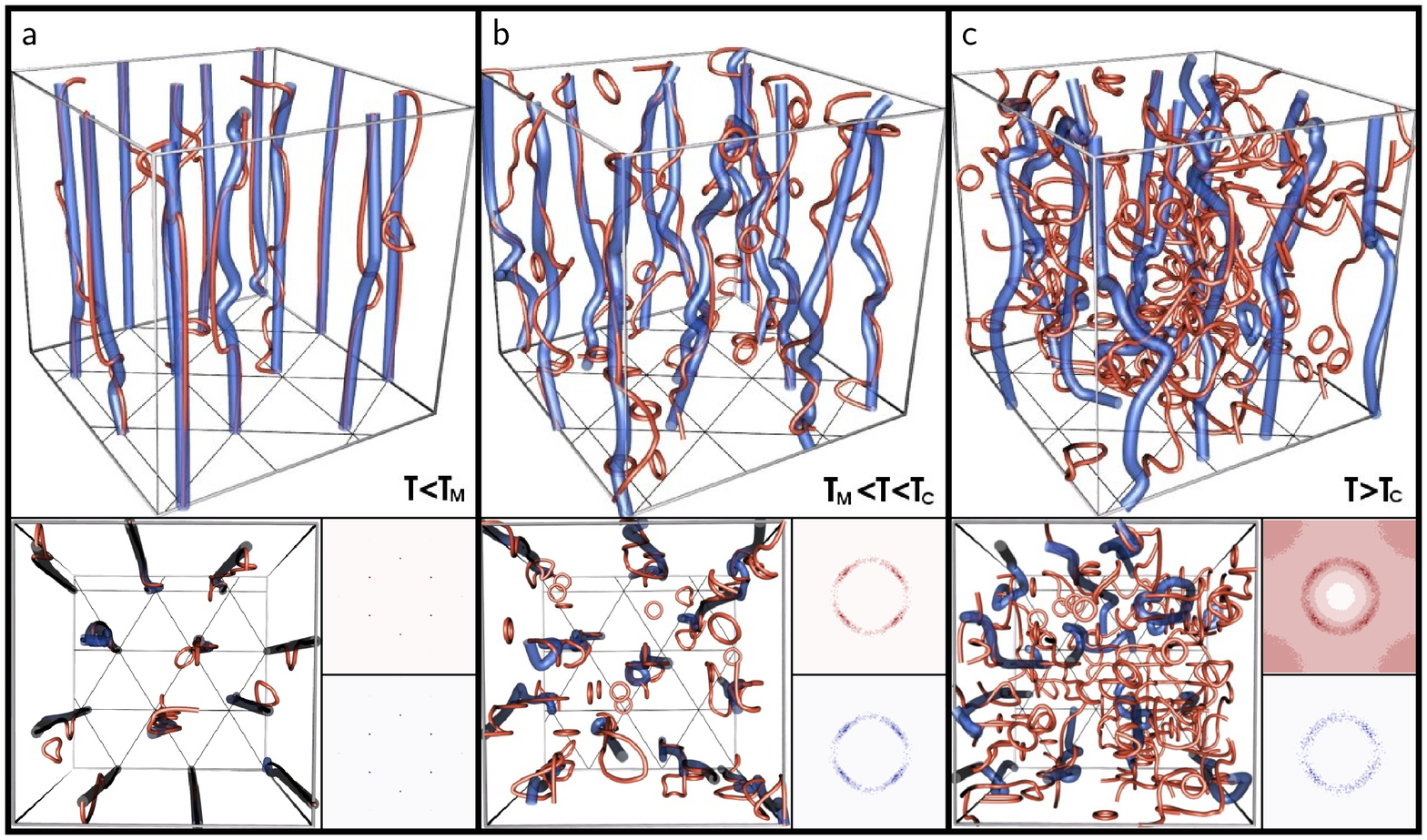}}}} 
        \caption{ \label{charge1} (Color online)
          Snapshots of the states of vortex
          matter and momentum space structure functions generated from
          MC simulations, taken at three different temperatures:
          $T=0.50 ~ (T<T_M)$, $T=0.72 ~ (T_M<T<T_C)$, and $T=0.86 ~ (T>T_C)$. 
          The snapshots are extracted from a small segment 
          ($15 \times 15 \times 15$) of the vortex system. Each thick frame 
          contains a sideview and a topview of the vortices, as well as 
          the protonic structure function $S^{(1)}(\ve k_{\perp})$ (red) and 
          the electronic structure function $S^{(2)}(\ve k_{\perp})$ (blue). 
          Panel {\bf a}: For $T<T_M$ the vortices are
          arranged in a co-centered lattice. Electronic (thick blue) and
          protonic(thin red) vortices only perform small excursions from
          each other. Both structure functions exhibit a sixfold Bragg
          pattern characteristic of a vortex lattice.  Panel {\bf b}:
          For $T_M < T < T_C$ the composite vortex lattice has melted as
          illustrated by the ring pattern in the structure functions.
          The electronic and protonic vortices perform stronger
          excursions from each other, but essentially remain co-centered.  
          This is the superfluid metallic phase in which co-directed 
          currents of protonic and electronic Cooper-pairs can propagate
          without dissipation. Panel {\bf c}: For $T>T_C$ the superfluidity
          is lost and the electronic and protonic vortices are no longer 
          co-centered. The proliferation of the protonic vortices is reflected 
          in the increase in the uniform background of the protonic structure 
          function.}
\end{figure}
\end{center}
\end{widetext}

Panel {\bf a} of Fig. \ref{charge1}  shows the side- and 
top-views of vortex matter at a small, but finite, temperature when two species of 
vortices perform only small excursion from each  other. The insets show 
the structure function in momentum space of the protonic (thin red) and electronic 
(thick blue) vortices. Parallel and co-directed vortices interact with each other as 
positively and negatively charged strings, and the splitting is a temperature-induced 
fluctuation. Panel {\bf b} shows snapshots of the vortex matter when the 
system is heated above the temperature of vortex lattice melting. The protonic 
(red) and electronic (blue) structure functions have developed ring-like structures, 
characteristic of a vortex-liquid in both the protonic and electronic sectors. 
While the vortices perform stronger excursions from each other, these excursions 
are still limited and one can always identify a red line attached to any given blue 
line. Thus, co-centricity of protonic (red)  and electronic (blue) vortex lines 
is still largely intact. In such a configuration, a dissipationless 
electrical current cannot propagate in any direction. Quite remarkably, however,  
co-directed currents of protonic and electronic pairs can 
propagate through this system without dissipation, as evidenced by the measurements of 
the helicity modulus given in Fig. \ref{Monte_Carlo_res}. Panel {\bf c} shows a 
state of vortex matter which occurs above the vortex ``ionization"  temperature. 
The co-centricity of vortices is strongly reduced, see Fig. \ref{Monte_Carlo_res}.  

This is also reflected in the subtle difference between protonic (red) and electronic 
(blue) structure functions in passing from panel {\bf b} to panel {\bf c}. {\it While the 
structure function of the electronic vortices essentially is unaffected by passing through 
the temperature $T_C$, the structure function for the protonic vortices is distinctly 
further isotropized.} Namely, the relative increase of the uniform background for the 
structure function of protonic vortices is a manifestation of the fact that protonic 
vortices suffer a vortex-loop proliferation transition inside the metallic vortex liquid 
phase. We have argued above that this transition belongs to the \xy universality class, 
i.e. the same universality class as the superfluid-to-normal fluid transition in liquid 
helium $^4$He \cite{Kleinert}.  

Concluding, we  report the observation, in a numerical experiment, of a novel dissipationless 
quantum state of matter, namely the {\it metallic superfluid}. Such a state might be realized 
in hydrogen or deuterium, if those systems were to take up a projected low-temperature 
{\it liquid metallic state} at an extreme pressure. The highest pressure obtained to date 
appears to be  around $320$GPa \cite{Loubeyre2002}. However, recent breakthroughs 
in artificial ultrahard diamond synthesis technology \cite{newlink} represent significant 
progress towards achieving extreme pressures in diamond anvil cells. 
Thus, a 
metallic superfluid might be the next ``super" state of matter to be realized in the 
laboratory.

This work was funded by the NTNU  through the Norwegian High Performance Computing Program, 
by the Research Council of Norway, Grant Nos. 157798/432, 158518/431, and 158547/431 (NANOMAT), 
by STINT and the Swedish Research Council, and NSF grant DMR-0302347. We thank N. W. Ashcroft 
for numerous discussions.

\end{document}